\documentclass[twocolumn,showpacs,preprintnumbers,amsmath,amssymb]{revtex4}

\usepackage{graphicx}
\usepackage{dcolumn}
\usepackage{bm}

\begin{document}

\preprint{APS/123-QED}

\title{Glass transition and alpha-relaxation dynamics of thin films of 
labeled polystyrene}

\author{Rodney D. Priestley}
\author{Linda J. Broadbelt}
\author{John M. Torkelson}
 \altaffiliation[]{Corresponding author: j-torkelson@northwestern.edu}
\affiliation{%
Department of Chemical and Biological Engineering
Northwestern University,
Evanston, IL 60208-3120, USA
}%

\author{Koji Fukao}
\altaffiliation[]{Corresponding author: kfukao@se.ritsumei.ac.jp}
\affiliation{%
Department of Macromolecular Science and Engineering, 
Kyoto Institute of Technology, 
Matsugasaki, Kyoto 606-8585, Japan
}%



\date{\today}

\begin{abstract}
The glass transition temperature and relaxation dynamics of the
segmental motions of thin films of polystyrene labeled with a dye,
4-[N-ethyl-N-(hydroxyethyl)]amino-4-nitraozobenzene (Disperse Red 1, 
DR1) are investigated using dielectric measurements.
The dielectric relaxation strength of the DR1-labeled polystyrene is 
approximately 65 times larger than that of 
the unlabeled polystyrene above the glass
transition, while there is almost no difference between them below the
glass transition. The glass transition temperature of the DR1-labeled
polystyrene can be determined as a crossover temperature at which 
the temperature coefficient of the electric capacitance changes from the
value of the glassy state to that of the liquid state.
The glass transition temperature of the DR1-labeled polystyrene decreases
with decreasing film thickness in a reasonably similar manner to that of the 
unlabeled polystyrene thin films.  The dielectric relaxation spectrum of the 
DR1-labeled polystyrene is also investigated. As thickness decreases, 
the $\alpha$-relaxation time becomes smaller and the distribution of the
$\alpha$-relaxation times becomes broader. These results show
 that thin films of DR1-labeled polystyrene are a suitable system
 for investigating confinement effects of the glass transition
 dynamics using dielectric relaxation spectroscopy.

\end{abstract}


\maketitle

\section{Introduction}
In recent years, intensive studies on the dynamics and the glass transition
in confined systems have been undertaken to elucidate the nature of the glass
transition~\cite{confinement2003,IDMRCS4}. 
The most promising scenario for the mechanism of the
glass transition is based on the Adam-Gibbs theory, in which a 
length scale characteristic of the dynamics associated with 
structural relaxation increases with decreasing temperature 
from the liquid state to the glassy state~\cite{Adam-Gibbs}. 
A major motivation for 
studies on the glass transition in a confined geometry was to measure
the characteristic length scale directly using different experimental 
techniques~\cite{Schuller,Kremer1}.

Polymer thin films are one of the ideal confined systems for such 
investigations because the system size, i.e., film thickness, can
be easily controlled experimentally. For this reason, many investigations
were conducted on thin polymer films with various film thickness  
to measure the glass transition temperature ($T_g$) and
the dynamics of the $\alpha$-process, which corresponds to the
structural relaxation and is related to the cooperative 
segmental motion of polymer
chains. For thin films supported on a substrate, many experimental results 
show that $T_g$ decreases with decreasing film thickness if there
is no strong attractive interaction, although there have been some 
conflicting experimental 
results~\cite{Keddie1,Keddie2,DeMaggio,Wallace,Jerome1,Efremov,Wubbenhorst1}.  
In particular, a very large decrease
in $T_g$ has been reported in freely-standing films of 
polystyrene (PS)~\cite{Forrest1,Forrest1a}.

The dynamics of thin polymer films have been investigated by many
experimental methods such as dynamic light scattering~\cite{Forrest2}, 
dielectric relaxation 
spectroscopy~\cite{Fukao1,Fukao2,Kremer2,Kremer3,Wubbenhorst2}, 
dynamic mechanical measurement~\cite{Tanaka1}, second harmonic
generation (SHG)~\cite{Torkelson2}, and so on. 
In accordance with the decrease in $T_g$, the dynamics of the
$\alpha$-process, which are directly associated with the glass
transition, become faster with decreasing film thickness.
In previous studies by Fukao et al., dielectric relaxation spectroscopy was 
applied to the investigation of the dynamics of ultrathin 
polymer films and provided much information about 
the relaxation dynamics of the $\alpha$-process, the
$\beta$-process and the normal mode in the case of 
polystyrene~\cite{Fukao1,Fukao2,Fukao3}, 
poly(vinyl acetate)~\cite{Fukao4}, poly(methyl
methacrylate)~\cite{Fukao4,Fukao5}, and cis-poly(isoprene)~\cite{Fukao5}.
Although the glass transition and dynamics of thin films of polystyrene 
have been investigated intensively, it is very difficult to obtain the
dielectric loss signal of the $\alpha$-process due to the very low
polarity of polystyrene.   

The above studies on the dynamics of thin polymer films are mainly 
related to the average $T_g$ and the average relaxation time of
the $\alpha$-process. However, it has been expected that there is a 
distribution or a positional dependence of $T_g$ and the relaxation time
of the $\alpha$-process within polymer thin films, especially thin supported
films. Ellison and Torkelson prepared multilayer films of labeled
and unlabeled polystyrene and successfully showed that there is a large
difference in $T_g$ between the regions near the free surface and  
near the substrate using fluorescence measurements~\cite{Torkelson3}. 
The $T_g$ of a 14-nm-thick layer at the free
surface is 32 K lower than the bulk $T_g$, while $T_g$ near the
substrate is equal to the bulk $T_g$. It may be expected that
analogous studies involving dielectric measurements of multilayer
polystyrene films can reveal the distributions of the dynamics 
within thin film layers, which are consistent with the distribution of $T_g$. 
However, such studies have not been previously conducted because no one had 
developed a system in which only one layer of a multilayer film is
dielectrically active.

As for the use of guest dipoles to enhance the dielectric response of
a weak polar material, there are several reports in the literature
~\cite{Williams1,Williams2,Williams3,Jager1,Jager2,Richert1,Ferreira1}. 
Thus, it has been previously established that the incorporation of guest 
dipoles into a non-polar polymer is a useful method to enhance its 
dielectric strength.   
In this study, we investigate the dielectric properties of single layer
films of polystyrene labeled with a nonlinear optical dye DR1 with various film
thicknesses and compare them with those previously 
observed for unlabeled polystyrene
in order to discuss the possibility of position-dependent measurements 
of the dynamics of the $\alpha$-process for a multilayer film.
It should be noted that this work is the first study to investigate the
impact of confinement using dielectric relaxation spectroscopy of a
labeled polymer.
 
This paper consists of six sections.
After giving experimental details in Sec.~II, the glass transition
temperature of thin films of PS labeled with DR1 is shown in Sec.~III. 
In Sec.~IV,
experimental results on the dielectric relaxation of the
$\alpha$-process of thin films of PS labeled with DR1 are given. 
After discussing the experimental results in Sec.~V, 
a summary of this paper is given in Sec.~VI.

\section{Experiments}
In the present study, we use polystyrene labeled at a low level with a 
nonlinear optical dye,
4-[N-ethyl-N-(hydroxyethyl)]amino-4-nitraozobenzene (Disperse Red 1,
DR1) (Fig.~1), which we refer to as PS-DR1.
The PS-DR1 is a random copolymer of neat styrene monomer and
DR1-labeled monomer synthesized following a procedure outlined
in Ref.~\cite{Torkelson4}. Hence, the dye molecules DR1 are 
covalently attached to the polymer chains of polystyrene. 
The concentration of DR1 in PS-DR1 is 
approximately 3.0 mol \% and $M_w$=1.34$\times$10$^4$ g/mol and $M_w/M_n$=1.65.
The molecular dipole moment of DR1 is approximately 7.0 D~\cite{Cheng1}.  
It has been established that when doped in PS, the DR1 reorientation 
dynamics measured by SHG
are coupled to cooperative segmental dynamics and can be used as a probe
of the $\alpha$-process~\cite{Torkelson1}. Also, a related study
demonstrated that the reorientation dynamics of DR1 
labeled to polymers are coupled to the $\alpha$-process~\cite{Torkelson4}.
The incorporation of DR1 from 0.0 to 3.0 mol \% label content to PS
results in a linear enhancement of the dielectric response, 
indicating that dye-dye associations (dipole quenching) do not occur in 
3.0 mol\% PS-DR1.

Thin polymer films are prepared by spin-coating from a toluene 
solution of PS-DR1 onto an aluminum(Al)-deposited glass substrate. 
Film thickness is controlled by changing the
concentration of the solution and spin speed of the spin-coater. 
The thin films obtained by spin-coating are annealed 
{\it in vacuo} for 48 hr at 303 K.  
After annealing, Al is vacuum-deposited onto the
thin films to serve as an upper electrode. 
Vacuum deposition of Al might increase the temperature of thin 
polymer films locally. However, no dewetting of the polymer films
are observed during the vacuum deposition of Al. Therefore, the
local heating of thin polymer films by vacuum deposition, if any, 
would not affect the present experimental results. 
The thickness of the Al electrode is controlled to be 40 nm, which is 
monitored by a quartz oscillator, and the effective area of the 
electrode $S$ is 8.0~mm$^2$.
The thickness $d$ of PS-DR1 is evaluated from the electric capacitance after
calibration with the absolute thickness measured by an atomic force microscope.

Dielectric measurements are performed using an LCR meter (HP4284A)
for the frequency range $f$ from 20~Hz to 1~MHz and an impedance analyzer
with a dielectric interface (Solartron Instruments 1260/1296) 
for the frequency range from 0.1~Hz to 1~MHz. The temperature of 
a sample cell is changed between 273~K and 413~K at a constant rate of
1~K/min. The dielectric measurements during the heating and cooling
processes are performed repeatedly several times.
Data acquisition is made during the above cycles except the first
cycle. Good reproducibility of dielectric data is obtained 
after the first cycle.

As shown in a previous study~\cite{Fukao2}, the resistance of the Al 
electrodes cannot be neglected for dielectric measurements of very 
thin films. This resistance leads to an artifact loss peak on the high 
frequency side.
Because the peak shape in the frequency domain is described by a Debye-type
equation, the ``C-R peak'' can easily be subtracted. Thus, the corrected
data are used for further analysis in the frequency domain.


\section{Glass transition temperature of thin films of PS labeled with DR1}

\begin{figure}
\includegraphics*[width=8cm]{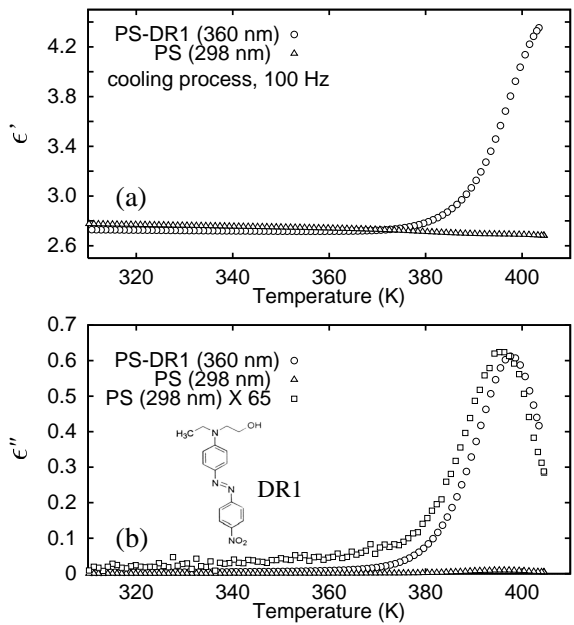}
\caption{\label{fig:1} Temperature dependence of the real 
and imaginary parts of the complex dielectric constants, $\epsilon'$
and $\epsilon''$, for PS-DR1 ($d$ = 360 nm) and unlabeled PS ($d$ = 298 nm). 
The data points in this figure are obtained at $f$ = 100 Hz during the
 cooling process. 
In Fig.~1(b), the values of $\epsilon''$ for the unlabeled PS 
magnified by 65 times are also plotted with square symbols for comparison.}
\end{figure}

Figure 1 shows the real and imaginary components of the
complex dielectric constant observed during the 
cooling process at the frequency of the applied electric 
field $f$~= 100 Hz for PS-DR1 and unlabeled PS films.
The thicknesses of the PS-DR1 and the unlabeled PS films are
360 nm and 298 nm, respectively, and hence both can be regarded as bulk
systems. It is clear that below $T_{\rm g}$
there is almost no difference in $\epsilon'$ and $\epsilon''$ 
between the PS-DR1 and the unlabeled PS.
However, the values of  $\epsilon'$ and $\epsilon''$ above $T_{\rm g}$
of PS-DR1 are much more enhanced compared to those of the unlabeled 
PS. As shown in Fig.~1(b), the peak height of the dielectric
loss due to the $\alpha$-process of PS-DR1 is approximately 
65 times larger than that of the unlabeled PS. 

\begin{figure}
\includegraphics*[width=8cm]{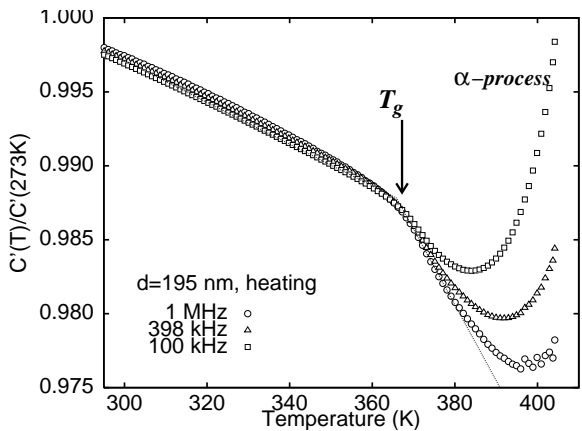}
\caption{\label{fig:2} Temperature dependence of the real part of
the electric capacitance normalized by the value at 273 K for
thin films of PS-DR1 with $d$ = 195 nm for three different frequencies
 100 kHz, 398 kHz, and 1 MHz. These data are observed during 
the heating process.}
\end{figure}

Figure 2 shows the temperature dependence of the real
part of the complex electric capacitance ($C'$) normalized to the value at
273~K for three different frequencies. The data are observed 
during the heating process. From this figure, it is observed that the 
normalized electric capacitances of the three different frequencies 
overlap and have a linear temperature dependence
below about 368~K. On the other hand, above this temperature
the electric capacitance has a linear temperature 
dependence with a larger slope 
but has a strong frequency dependence at even higher temperatures. 
As a result, the electric capacitance
deviates from the common straight line. As previously discussed, 
the negative slope of the straight line $\tilde\alpha$ corresponds to 
the thermal expansion coefficient normal to the film 
surface $\alpha_n$~\cite{Fukao1,Fukao2}.
If the lateral size of the sample does not change, we have the following
relation $\tilde\alpha\approx 2\alpha_n$. Judging from the electric 
capacitance in Fig.~1, the value of $\alpha_n$ increases drastically at
368~K. Therefore, this temperature can be regarded as 
the $T_{\rm g}$ of PS-DR1 with $d$~=~195 nm. The linear 
thermal expansion coefficient $\alpha_n$ evaluated from Fig.~2 changes
from 0.7$\times$10$^{-4}$~K$^{-1}$ to 2.5$\times$10$^{-4}$~K$^{-1}$, 
which agrees well with 
the literature values of PS. The frequency dispersion of $C'$ 
above $T_{\rm g}$ is due to the $\alpha$-relaxation
process.

\begin{figure}
\includegraphics*[width=8cm]{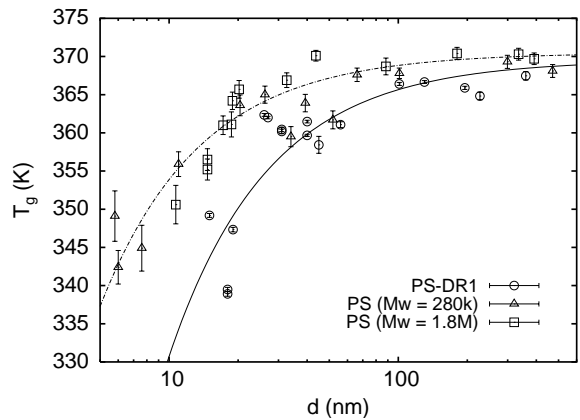}
\vspace*{-0.2cm}
\caption{\label{fig:3} Thickness dependence of $T_{\rm g}$ determined
from the crossover temperature as shown in Fig.~2 for thin films of
 PS-DR1. The thickness dependence of $T_{\rm g}$ of the unlabeled PS
with $M_w$ = 2.8$\times$10$^5$ and 1.8$\times$10$^6$ g/mol is also 
plotted~\cite{Fukao1,Fukao2}. The solid and dashed curves are obtained
 by fitting the observed data points to Eq.~(1) in the text.}
\end{figure}

Figure 3 shows the thickness dependence of $T_{\rm g}$ determined 
from the temperature dependence of the electric capacitance as mentioned above.
This figure clearly shows that $T_{\rm g}$ decreases with decreasing
film thickness. The thickness dependence of $T_{\rm g}$ can be fitted
using the following equation:
\begin{eqnarray}
T_{\rm g}(d) &=&T_{\rm g}^{\infty}\left(1-\frac{a}{d}\right),
\end{eqnarray}
where the best fit parameters are $T_{\rm g}^{\infty}$ = 369.5 $\pm$ 2.1 K 
and $a$ = 1.1 $\pm$ 0.2 nm. 
In Fig.~3 the thickness dependence of $T_{\rm g}$ of 
unlabeled PS is also shown. The molecular weights of the unlabeled PS
are $M_w$ = 2.8$\times$10$^5$ and 1.8$\times$10$^6$ g/mol. The values of 
$T_{\rm g}$ of the unlabeled PS are determined using the same method 
as previously reported~\cite{Fukao1}. Comparing the thickness dependence of
$T_{\rm g}$ between the PS-DR1 and the unlabeled PS, it is found that
in both cases $T_{\rm g}$ decreases with decreasing film thickness 
in a reasonably similar way. At the same time, we notice that there is 
a small difference between the two cases. This difference may be
attributed to the difference in molecular weight between the PS-DR1 and
the unlabeled PS, because the molecular weight of PS-DR1 
($M_w$ = 1.34$\times$10$^4$) is much smaller than that of the unlabeled 
PS, and accordingly $T_{\rm g}$ of the bulk state of PS-DR1 is
lower by a few degrees than that corresponding to the unlabeled PS.
If we take into account the difference coming from the molecular weight,
we can judge that the thickness dependence of $T_{\rm g}$ of PS-DR1
is fairly comparable to that of the unlabeled PS.

The use of dielectric spectroscopy requires that an upper electrode be
evaporated on top of the free surface of the films.  A consequence is
that the films do not have a true free surface.  However, the PS films
report a reduction in $T_{\rm g}$ with decreasing film thickness similar 
to techniques that allow for a free surface.  Therefore, it is believed 
that the free surface effects are not masked by the addition of the top 
electrode as demonstrated previously~\cite{Fukao1,Kremer2,Wubbenhorst2}.

\section{Dynamics of the $\alpha$-process in thin films of PS labeled with DR1}
\subsection{Dielectric behavior at a fixed frequency}
\begin{figure}
\includegraphics*[width=8cm]{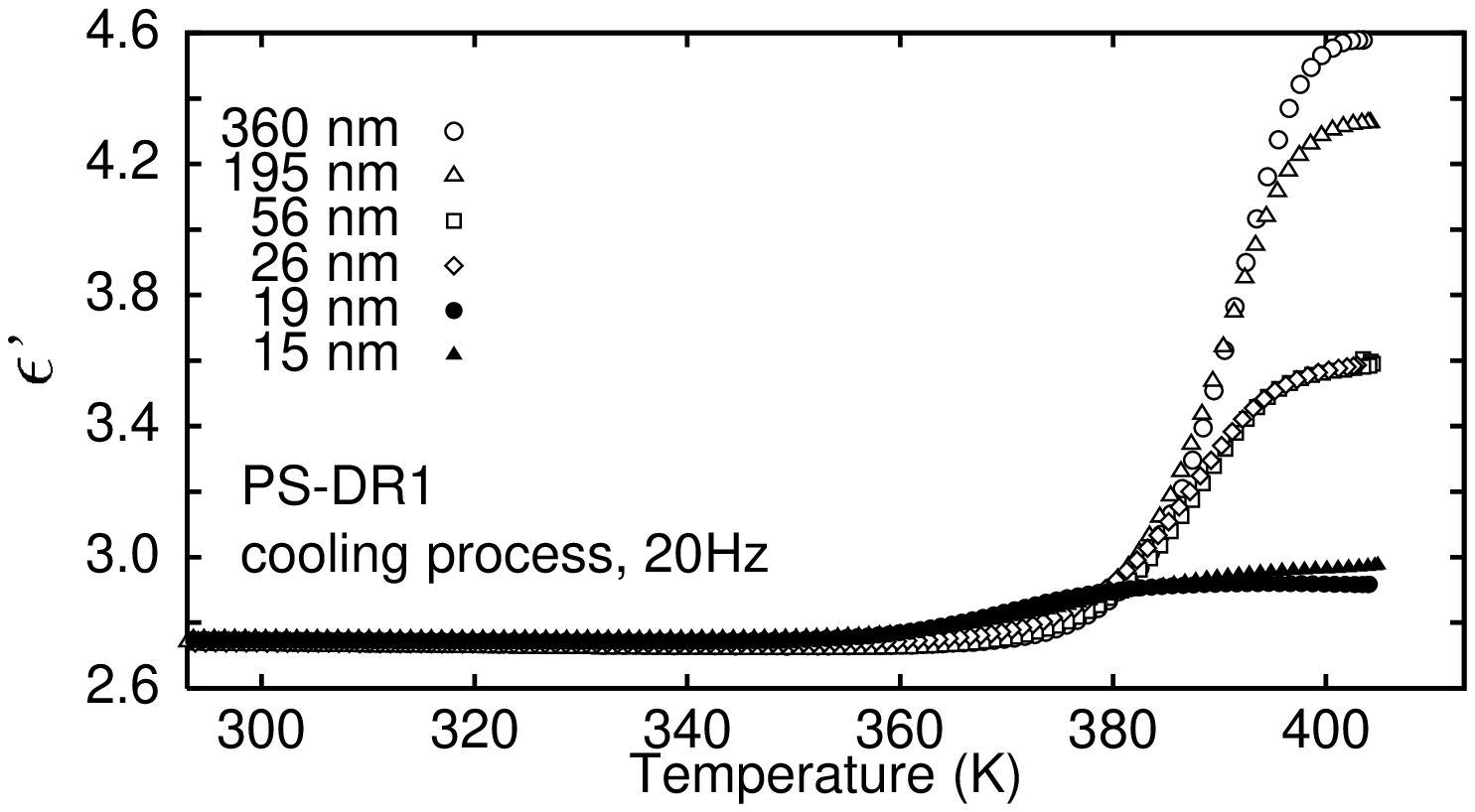}
\includegraphics*[width=8cm]{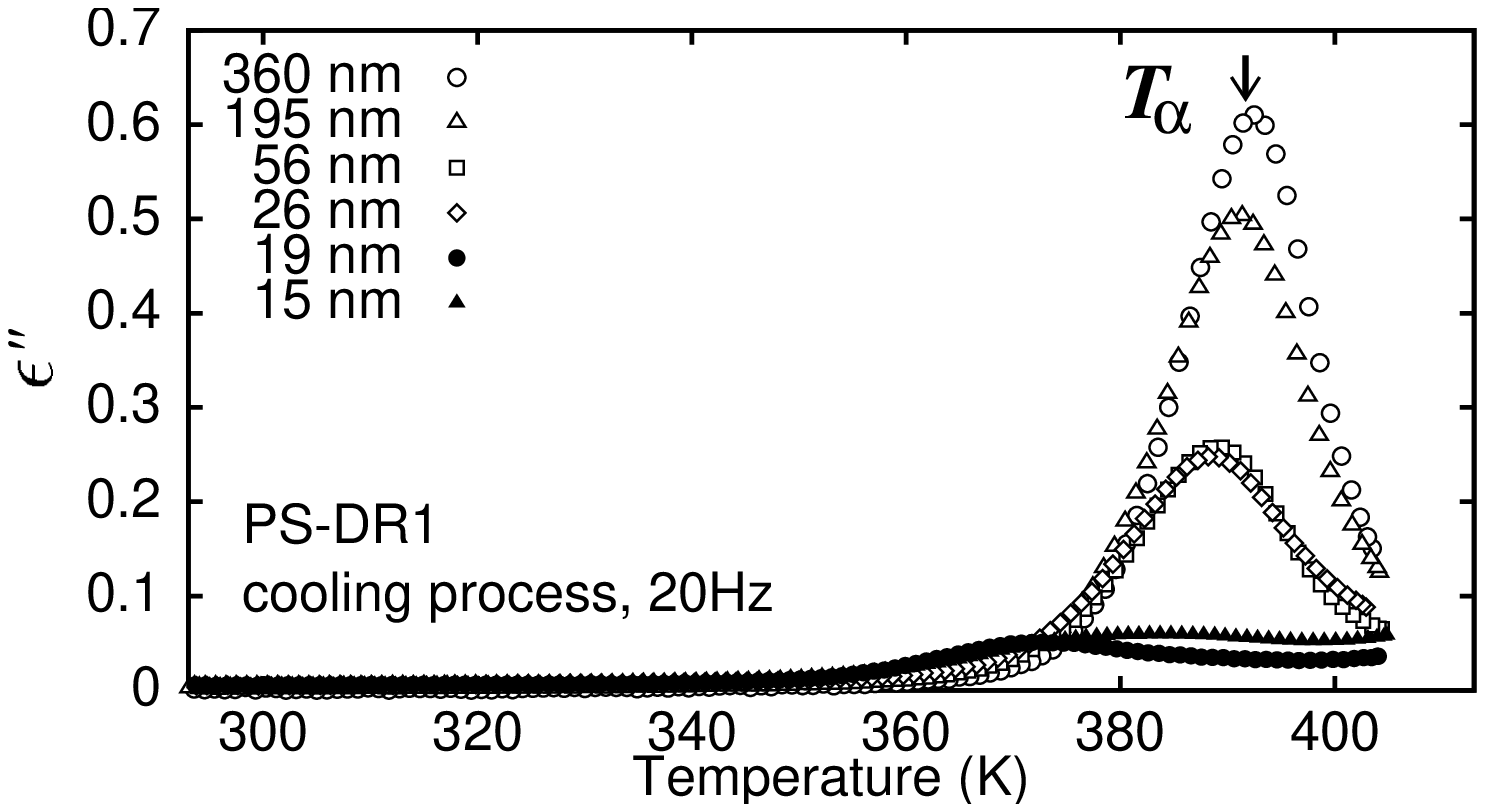}
\caption{\label{fig:4} Temperature dependence of the real and imaginary
 components of the dielectric constants $\epsilon'$ and $\epsilon''$ at 
the frequency 20 Hz for different film thicknesses.} 
\end{figure}

\begin{figure}
\includegraphics*[width=8cm]{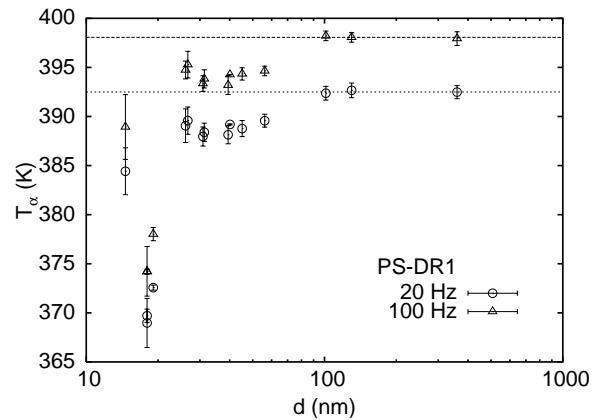}
\caption{\label{fig:5} Thickness dependence of the temperature 
$T_{\alpha}$ observed at 20~Hz and 100~Hz. 
The dotted and dashed lines correspond to
 the values observed in bulk states at 20~Hz and 100~Hz, respectively.}
\end{figure}

Figure 4 shows the temperature dependence of the real and imaginary
components of the complex dielectric constant for thin films of PS-DR1
with film thickness ranging from 360~nm to 15~nm. The data are obtained
during the cooling process at frequency 20 Hz. In Fig.~4 we find that 
the contribution from the $\alpha$-process is strongly affected by
film thickness. As the thickness decreases, the peak height
of the dielectric loss due to the $\alpha$-process becomes
smaller, and at the same time the $\alpha$-temperature ($T_{\alpha}$)
at which the dielectric loss due to the $\alpha$-process has
a maximum is shifted to the lower temperature side. In Fig.~5, 
$T_{\alpha}$ is plotted as a function of film thickness for $f$~=~20~Hz and
100~Hz. The temperature $T_{\alpha}$ is found to depend strongly 
on frequency and to increase with increasing frequency. 
The value of $T_{\alpha}$ at a low frequency 
corresponding to the $\alpha$-relaxation time of 100~sec 
is comparable to the value of $T_{\rm g}$ determined for the
ramping process at a rate of 10~K/min using 
dilatometric measurements or differential scanning calorimetry.
Therefore, the decrease in $T_{\alpha}$ with decreasing film 
thickness at a given
frequency is associated with the decrease in $T_{\rm g}$ and the
faster dynamics of the $\alpha$-process in thinner films.

\subsection{Dielectric relaxation in thin films}

%
%
\begin{figure*}
\includegraphics*[width=8cm]{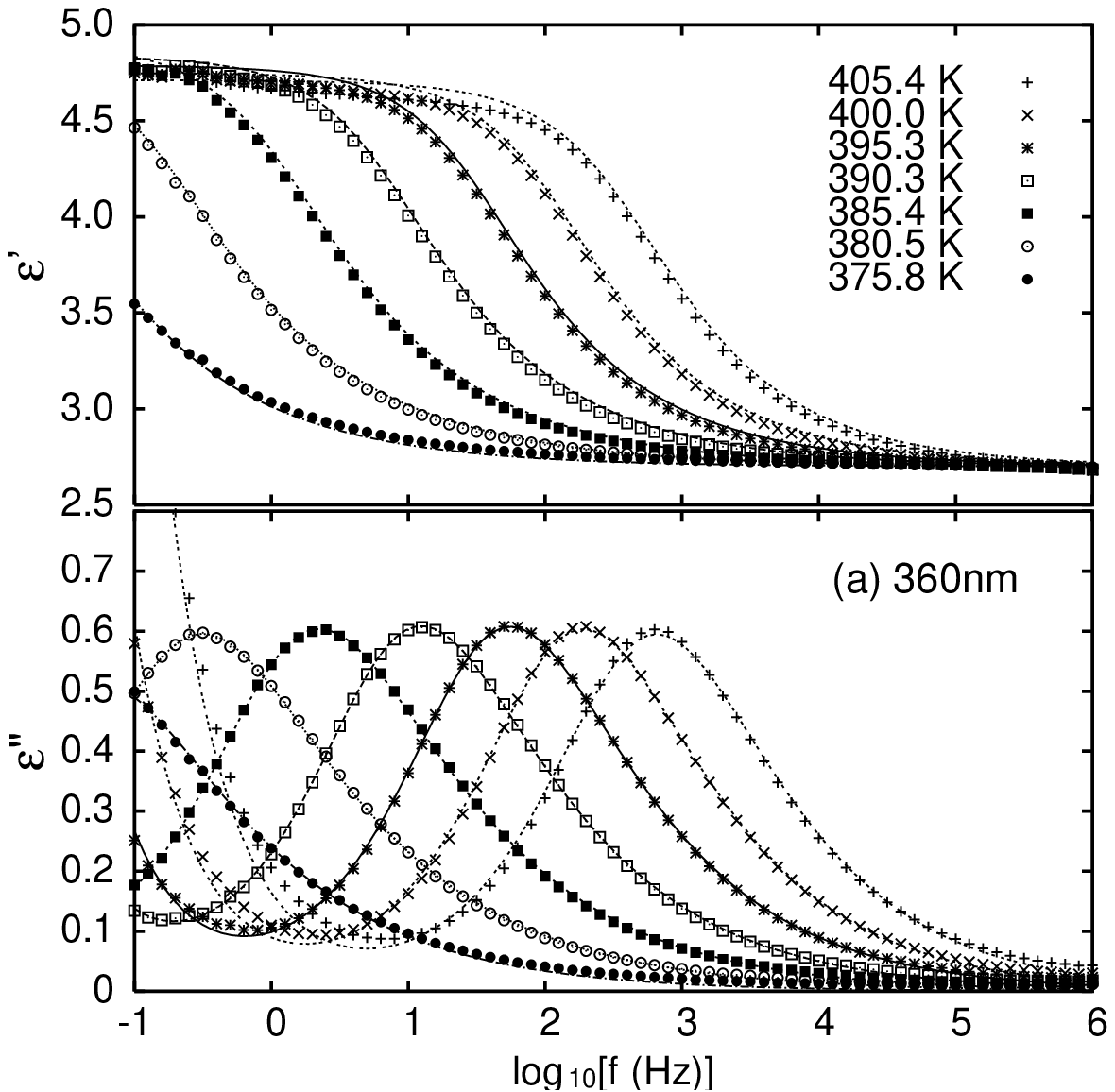}
\includegraphics*[width=8cm]{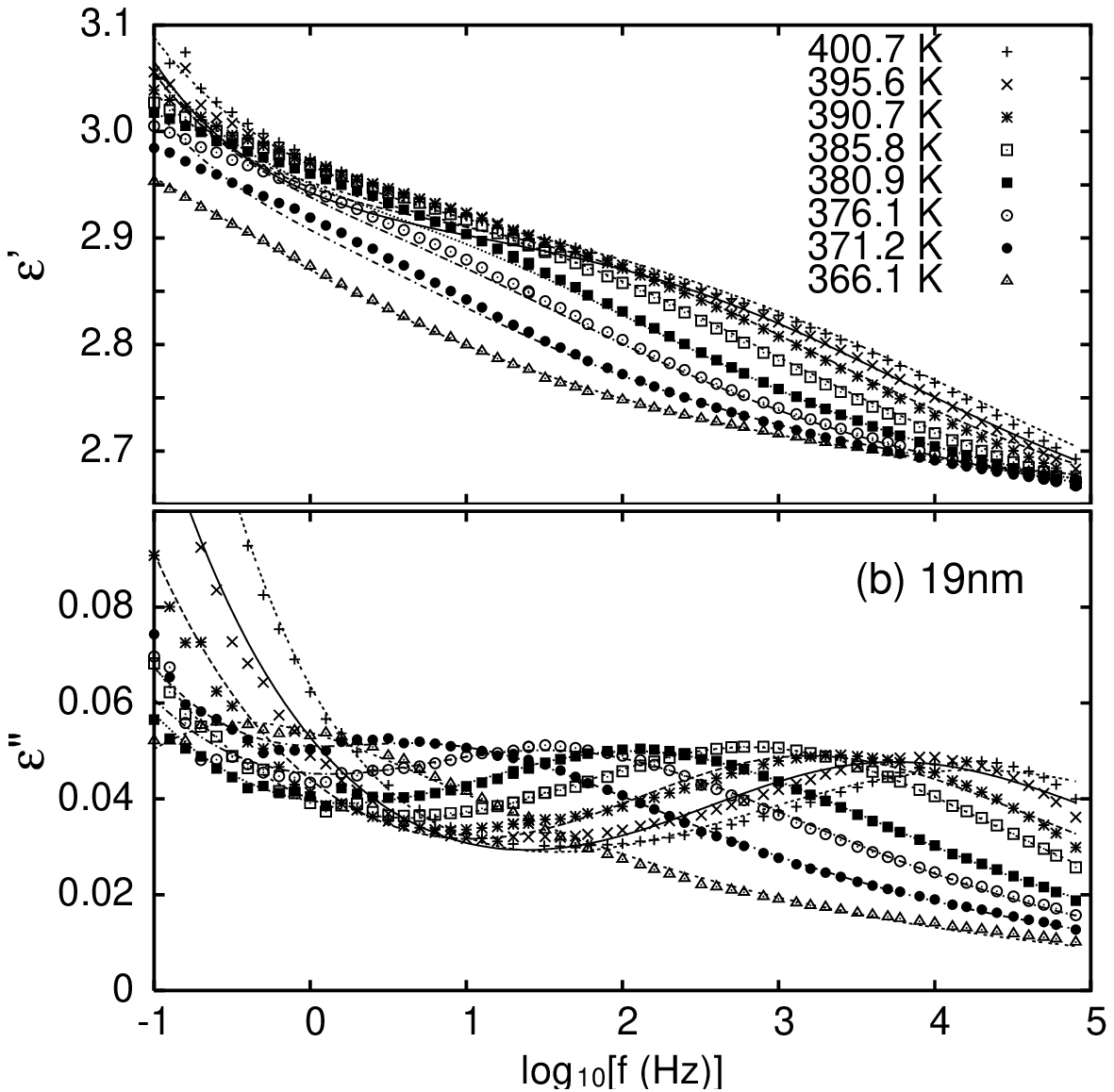}
\caption{\label{fig:6} The dependence of the complex dielectric constant
 on the logarithm of frequency at various temperatures above $T_g$ 
for thin films of
PS-DR1: (a) $d$ = 360 nm and (b) $d$ = 19 nm. Solid curves are
 calculated by Eq.~(2).}
\end{figure*}

\begin{table*}
\caption{Fitting parameters of the HN equation for PS-DR1 with various
 film thicknesses: relaxation strength $\Delta\epsilon$, 
shape parameters $\alpha_{HN}$, $\beta_{HN}$, and
relaxation time $\tau_0$.
Here, the exponent $\beta_{KWW}$ in the KWW relaxation function 
is evaluated using the relation 
$\beta_{KWW}=(\alpha_{HN}\beta_{HN})^{1/1.23}$~\cite{colmenero}.}
\begin{tabular}{ccccccc}\hline
$d$ (nm) & $T$ (K) & $\Delta\epsilon$ & $\alpha_{HN}$ & $\beta_{HN}$ &
 $\tau_0$ (sec) & $\beta_{KWW}$
\\\hline
19 & 390.7 & 0.33 $\pm$ 0.02 & 0.46 $\pm$ 0.02 & 0.48 $\pm$ 0.10 &
(3.1$\pm$0.5)$\times$10$^{-4}$ & 0.29$\pm$0.06\\
26 & 389.8 & 0.96 $\pm$ 0.01 & 0.69 $\pm$ 0.02 & 0.52 $\pm$  0.02
 & (1.8$\pm$0.1)$\times$10$^{-2}$ & 0.43$\pm$0.24\\
360 & 390.3 & 2.14 $\pm$ 0.01 & 0.80 $\pm$ 0.01 & 0.57 $\pm$ 0.01 &
(2.32$\pm$0.04)$\times$10$^{-2}$ & 0.53$\pm$0.08 \\\hline
%
\end{tabular} 
\end{table*}

\begin{table}
\caption{The values of the parameters resulting in the best fit of the 
relaxation times of the $\alpha$-process $\tau$ to Eq.~(3) for thin films
 of PS-DR1 with various film thicknesses ($d$ = 360 nm, 26 nm and 19 nm).}
\begin{tabular}{ccccc}
\hline\hline
d (nm) & $\log_{10}[\tilde\tau_0$(sec)] & $U$ (10$^3$ K) & $T_0$ (K) & m \\\hline
360 & $-$11.8 $\pm$ 0.6 & 1.6 $\pm$ 0.2 & 318 $\pm$ 4 & 99 $\pm$ 11 \\
26 & $-$13.3 $\pm$ 0.5 & 2.1 $\pm$ 0.2  & 306 $\pm$ 3 & 92 $\pm$ 8\\
19 & $-$14.6 $\pm$ 1.0 & 2.3 $\pm$ 0.4 & 294 $\pm$ 7 & 97 $\pm$ 17\\
\hline\hline
\end{tabular}
\end{table}

Figure 6 shows the frequency dependence of the real and imaginary
components of the complex dielectric constants at various temperatures
for thin films of PS-DR1 for two different thicknesses: 
(a) $d$ = 360 nm and (b) $d$ = 19 nm. In the real part of the complex 
dielectric constant $\epsilon'$ for the 360nm-thick-film, 
there is a gradual step from
4.7 to 2.7, while in the imaginary part $\epsilon''$ there is 
a maximum at the same frequency where there is the step in $\epsilon'$.
This frequency dependence is associated with the existence of
the $\alpha$-process. The peak frequency in $\epsilon''$ 
corresponds to the inverse of 
a characteristic time of the $\alpha$-process at a given temperature.  
From Fig.~6 it is observed that the peak frequency of the $\alpha$-process
becomes larger 
with increasing temperature.
This corresponds to the acceleration of the dynamics of the
$\alpha$-process with increasing temperature. At higher temperatures 
there is also a large increase in $\epsilon''$ with decreasing 
frequency. This is usually attributed to the contributions
from dc conductivity due to space charges or impurities within the 
polymeric systems. 
Comparing the frequency dependences of 
$\epsilon'$ and $\epsilon''$ in Fig.~6(a)
to those in Fig.~6(b), we find that the peak height of the
loss peak in $\epsilon''$ due to the $\alpha$-process for $d$~= 19 nm
is much smaller than that for $d$ = 360 nm and that the peak shape and 
the peak position also change with decreasing film thickness.

Here, we use the following empirical equation 
of $\epsilon'$ and $\epsilon''$ as a function of frequency:
\begin{eqnarray}\label{HN}
\epsilon^*=\epsilon_{\infty}+i\frac{\tilde\sigma}{\epsilon_0}\omega^{-m}
+\frac{\Delta\epsilon}{[1+(i\omega\tau_0)^{\alpha_{HN}}]^{\beta_{HN}}},
\end{eqnarray}
where $\omega$ = $2\pi f$, $\epsilon_0$ is the permittivity {\it in vacuo}
and $\epsilon_{\infty}$ is the permittivity at
a very high frequency. The second term is a contribution from space
charge~\cite{Miyamoto}, 
and this contribution can be attributed to pure dc conductivity
if $m$ = 1. The third term comes from the $\alpha$-process, and its 
empirical form is usually called the Havriliak-Negami (HN) equation, where
$\Delta\epsilon$ is the relaxation strength, $\alpha_{HN}$ and
$\beta_{HN}$ are the shape parameters, and $\tau_0$ is the relaxation 
time of the $\alpha$-process. The solid curves in Fig.~6 are obtained
using Eq.~(\ref{HN}) with the best-fit parameters. In Fig.~6 it is
found that Eq.~(\ref{HN}) can well reproduce the frequency dependence of 
the observed dielectric constant.  Examples of the best-fit
parameters of the HN-equation at 390~K are listed in Table~I.
In this table, the exponent $\beta_{KWW}$ is also listed, on the
assumption that the relaxation function $\phi(t)$ is given by the KWW
equation $\phi(t)=\exp(-(t/\tau)^{\beta_{KWW}})$. The value of
$\beta_{KWW}$ can be evaluated by the relation 
$\beta_{KWW}=(\alpha_{HN}\beta_{HN})^{1/1.23}$ and is a measure of the
distribution of the relaxation times~\cite{colmenero}.

\subsection{Relaxation time of the $\alpha$-process}

\begin{figure}
\includegraphics*[width=8cm]{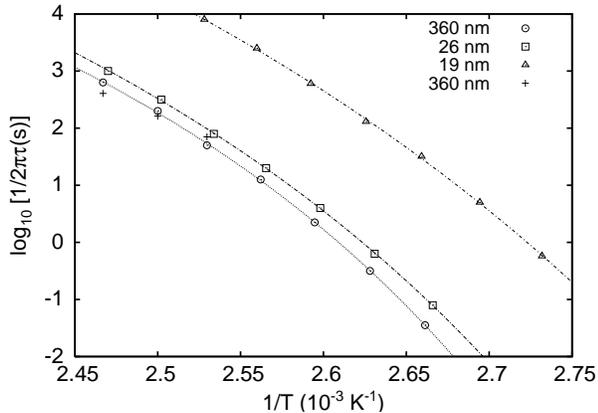}
\caption{\label{fig:7} Arrhenius plot for the $\alpha$-relaxation
 process in thin films of PS-DR1. 
The logarithm of 1/2$\pi\tau$ vs. $1/T$ for three different thicknesses
$d$ = 19 nm ($\protect\triangle$), 26 nm ($\protect\Box$), and 360 nm 
($\protect\circ$). The curves are obtained using the VFT law. The symbol
 $+$ corresponds to dc-conductivities which are evaluated at 0.1~Hz  
for $d=$~360~nm in Fig.10 and are shifted along the vertical axis 
so that we can compare the temperature dependence of dc-conductivity 
to that of  $1/\tau$ .}
\end{figure}

Figure 7 shows the Arrhenius plot of the $\alpha$-process of thin films
of PS-DR1 with $d$ = 19 nm, 26 nm, and 360 nm. The vertical axis is the
logarithm of 1/2$\pi\tau$, where $\tau$ is the relaxation time of the 
$\alpha$-process and is evaluated from the relation
$2\pi f_{\rm max}\tau$ = 1, where $f_{\rm max}$ is the frequency at which
$\epsilon''$ has a loss peak due to the $\alpha$-process at 
a given temperature.
The curves are evaluated using the Vogel-Fulcher-Tammann (VFT) law:
\begin{eqnarray}
\tau(T)=\tilde\tau_0\exp\left(\frac{U}{T-T_0}\right),
\end{eqnarray}
where $\tilde\tau_0$ is a microscopic time scale for the $\alpha$-process,
$U$ is an apparent activation energy, and $T_0$ is the Vogel
temperature~\cite{VFT}. 
For each film thickness it is found that the relaxation 
time of the
$\alpha$-process obeys the VFT law.
At the same time, there is a distinct
thickness dependence of $\tau$, that is, the relaxation time of the 
$\alpha$-process becomes smaller with decreasing film thickness at a
given temperature.  The best-fit parameters of the VFT law for thin 
films of PS-DR1 are listed in Table~II. It is clear that
the Vogel temperature decreases with decreasing film thickness, which
is consistent with the fact that $T_{\rm g}$ decreases with decreasing
film thickness as shown in Fig.~3. The fragility index $m$,
which is a measure of the non-Arrhenius temperature dependence
of the relaxation times, is also
evaluated from the temperature dependence of the $\alpha$-relaxation
time according to the following definition:
\begin{eqnarray}
m&=&\left[\frac{d\log_{10}\tau(T)}{d(T_{\rm g}/T)}\right]_{T=T_{\rm g}},
\end{eqnarray}
where $T_{\rm g}$ is defined so that $\tau(T_{\rm g})$ = 100~sec~\cite{Bohmer1}.

\begin{figure}
\includegraphics*[width=8cm]{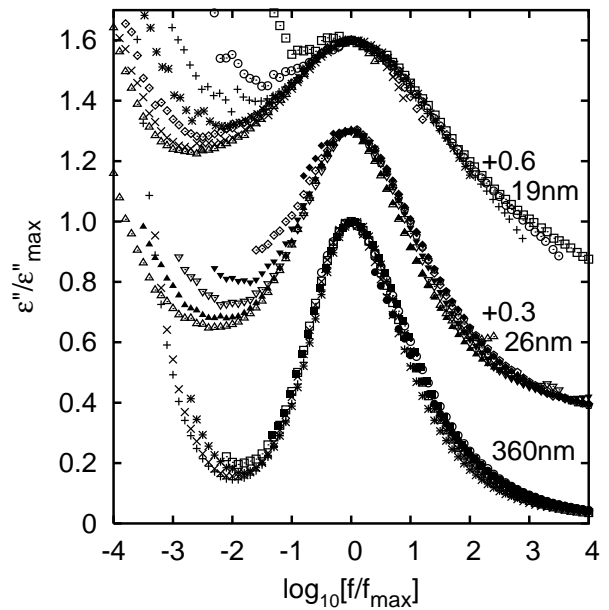}
\caption{\label{fig:8} Dependence of the normalized dielectric loss of
 PS-DR1 on
 the logarithm of the normalized frequency. The two axes are normalized
 with respect to the peak position due to the $\alpha$-process,
 corresponding to $\epsilon''_{\rm max}$ and $f_{\rm max}$. The numbers
 given in the right margin stand for the film thickness. The data for
 $d$ = 19 nm and 26 nm are shifted up by +0.6 and +0.3,
 respectively, for clarity. Different symbols correspond to 
different temperatures:
(a) for $d$ = 360 nm, data points at 405.4 K, 400.0 K, 395.3 K, 390.3 K, 385.4
 K, 380.5 K, and 375.8 K are plotted 
(b) for $d$ = 26 nm,  404.9 K, 399.7 K, 394.7 K, 
389.8 K, 384.9 K, 380.1 K, and 375.1 K (c) for $d$ = 19 nm,  
400.7 K, 395.6 K, 390.7 K, 385.8 K, 380.9 K, 376.1 K, and 371.2 K}
\end{figure}

\subsection{Profile of the $\alpha$-loss peak}

\begin{figure}
\includegraphics*[width=8cm]{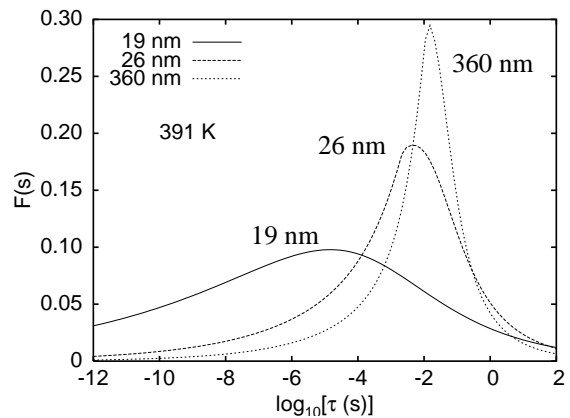}
\caption{\label{fig:9} The distribution function $F(s)$ of the
 relaxation times of the $\alpha$-process at 391 K for thin films of
 PS-DR1. The values are calculated by Eq.~(6) with the best fit
 parameters of the HN equation. The dotted curve represents the result for
$d$ = 360 nm, the dashed curve that for $d$ = 26nm, and the solid curve 
 that for $d$ = 19 nm.}
\end{figure}

In order to obtain the thickness dependence of the profile of the dielectric
loss spectrum, the observed loss peaks of $\epsilon''$ at various
temperatures are normalized with 
respect to the peak position for each temperature in the case of
three different film thicknesses, as shown in Fig.~8. 
For clarity, data points of $d$ = 19 nm and $d$ = 26 nm are shifted along the
vertical axis by +0.6 and +0.3, respectively.
From Fig.~8 it is found that the width of the $\alpha$-loss peak
clearly increases with decreasing film thickness. This suggests that
the distribution of the relaxation times becomes broader with decreasing
film thickness. At the same time, there is a contribution due to the
dc conductivity in the low-frequency side, which may disturb the
evaluation of the distribution of the relaxation times of the
$\alpha$-process. In order to avoid this problem, we use the
best-fit parameters of the HN equations $\alpha_{HN}$, $\beta_{HN}$
and $\tau_0$ and then evaluate the distributions of $\tau$.

Here, the distribution function $F(\log_e\tau)$ of the relaxation 
times $\tau$ is defined by the following relation:
\begin{eqnarray}
\epsilon^*(\omega)=\epsilon_{\infty}+\Delta\epsilon\int^{+\infty}_{-\infty}
\frac{F(\log_e\tau)d(\log_e\tau)}{1+i\omega\tau}.
\end{eqnarray}
If we assume that the shape of the dielectric loss peak
is described by the HN equation, the distribution function
$F(\log_e\tau)$ can be calculated analytically as follows:
\begin{eqnarray}
 F(s)&=&\frac{1}{\pi}[1+2e^{\alpha_{HN}(x_0-s)}\cos\pi\alpha_{HN}+e^{2\alpha_{HN}(x_0-s)}]^{-\beta_{HN}/2}\nonumber\\
&&\times\sin\left[\beta_{HN}\tan^{-1}\left(\frac{e^{\alpha_{HN}(x_0-s)}\sin\pi\alpha_{HN}}{1+e^{\alpha_{HN}(x_0-s)}\cos\pi\alpha_{HN}}\right)\right],
\end{eqnarray}
where $s=\log_e\tau$~\cite{Fukao4} ans $x_0=\log_e\tau_0$.
Figure 9 shows the distribution of $\alpha$-relaxation times for
three different film thicknesses at 391~K, which is evaluated using
Eq. (6). It is found that 
the relaxation time of the $\alpha$-process, which is related to 
the peak position of the distribution, is shifted
to a smaller time  with decreasing film thickness, and at the
same time, the full width at the half maximum of the distribution 
becomes broader, increasing from
2 decades (for $d$ = 360 nm) to 8 decades (for $d$ = 19 nm). 


\subsection{Conductivity component}

\begin{figure*}
\includegraphics*[width=16cm]{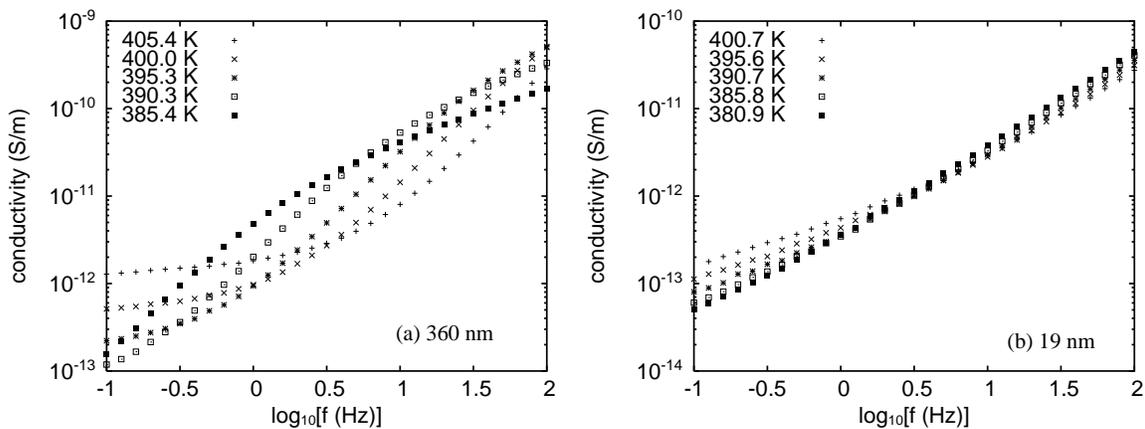}
\caption{\label{fig:10} The frequency dependence of the real part of 
the conductivity at various temperatures for PS-DR1 with $d=$~360~nm (a)
 and  19~nm (b).
The value of the vertical axis is evaluated from the frequency
 dependence of the imaginary part of the dielectric ac-susceptibility 
using the relation $\sigma' = \omega\epsilon_0\epsilon''$.
}
\end{figure*}
 
Figure 6 shows that there is a contribution of conductivity 
due to the motion of space charge such as ions included in polymer
materials in the low frequency range. In order to analyze this
contribution in the low frequency region, 
we use the second term of the right-hand side of Eq.~(2): 
$\epsilon''_{\rm con} \sim \frac{\tilde\sigma}{\epsilon_0}\omega^{-m}$. 
Making data fit to this equation for various temperatures, 
we can obtain the temperature dependence
of $\tilde\sigma$ and $m$ for $d=$ 19~nm and 360~nm. 
It is found that $m$ is approximately independent of temperature
and is equal to 0.82 $\pm$ 0.02 for $d=$ 360~nm and 0.48~$\pm$~0.01 for 
$d$=~19~nm.
If $\epsilon''_{\rm con}$ has a $\omega$ dependence given by $\omega^{-m}$, 
the real part of ac-conductivity $\sigma'$ is proportional
to $\omega^{1-m}$. Hence, we obtain the $\omega$ dependence of
$\sigma$ in the low frequency region as follows:
\begin{eqnarray}\label{t_alpha}
\sigma' \sim \left\{ \begin{array}{r@{\quad:\quad}l} 
\omega^{0.18} & d = 360 \mbox{~nm} \\
\omega^{0.53} & d = 19 \mbox{~nm}.
\end{array}\right.
\end{eqnarray}

In order to extract the dc-conductivity, the dielectric loss
$\epsilon''$ in Fig.~6 are replotted as $\omega\epsilon_0\epsilon''$
vs. $\log f$, where $\omega\epsilon_0\epsilon''$ corresponds to 
the real part of conductivity $\sigma'$, as shown in Fig.~10.
In Fig.~10, it is found that $\sigma'$ obeys the power-law $\omega^{1-m}$
in the lower frequency region and tends to approach a constant value
with decreasing frequency.
This constant value corresponds to the dc-conductivity. For $d=$~360 nm,
the value $\sigma'$ at 0.1~Hz in Fig.~10 can be regarded as the
dc-conductivity because the slope in the low frequency region is small. 
On the other hand, for $d=$~19~nm, $\sigma'$ still decreases
with decreasing frequency in the low frequency region around 0.1~Hz, and
hence it is impossible to evaluate the dc-conductivity from the present
data for $d=$~19~nm. The dc-conductivities obtained thus for $d=$~360~nm 
are plotted in Fig.~6 after shifting them along the vertical axis 
so that we can compare the temperature dependence of dc-conductivity
to that of $1/\tau$. In Fig.~6 it found that there is a fairly 
good agreement between the two data. Therefore, the present results are 
consistent with previous results that dc-conductivity 
has a similar temperature dependence of the segmental motion~\cite{Floudas1}.
For detailed comparison, the data at much lower frequencies are highly
required.



\section{Discussion}


In 1994, Torkelson and coworkers investigated the rotational dynamics of 
DR1 doped at 2~wt.\% in polystyrene using 
SHG and dielectric relaxation 
spectroscopy~\cite{Torkelson1}.
They found that both SHG and dielectric relaxation
spectroscopy yielded almost the same average time constant 
$\langle\tau\rangle$,
and that above $T_{\rm g}$ the values of $\langle\tau\rangle$ fit
well to the Williams-Landel-Ferry (WLF) equation with appropriate WLF 
constants~\cite{WLF}, 
which indicated that the rotational reorientation dynamics of DR1 are 
coupled to the $\alpha$-relaxation process of PS.

As shown in the previous section, the dielectric loss peak can be observed 
above $T_{\rm g}$ using dielectric relaxation spectroscopy for thin
films of PS-DR1. In our measurement, we observe a very large 
dielectric loss above $T_{\rm g}$ compared to the unlabeled PS.
In PS-DR1, DR1 chromophores are attached covalently to the main polymer
chain, while DR1 dyes were doped in PS in Ref.~\cite{Torkelson1}. 
Although there 
are covalent bonds between DR1 and PS in the present case, the
rotational reorientation 
relaxation times of the labeled DR1 can still be described 
by the VFT law, which is the same as the WLF equation, as shown in
Fig.~7.
This is consistent with results in Ref.~\cite{Torkelson1}. 
Therefore, the rotational 
reorientation dynamics of DR1 chromophores attached to the 
polymer main chain, which must be the microscopic origin of the dielectric 
loss observed in the present study, are equivalent to the cooperative segmental
motions of PS, that is, the $\alpha$-relaxation process.

Above $T_{\rm g}$, the reorientation dynamics of DR1 coupled with 
the $\alpha$-process have a very large contribution to the dielectric
susceptibility, as shown in Fig.~1.  In many polymeric systems with 
large polarity such as PMMA and PVAc, it is impossible to determine 
$T_{\rm g}$ using capacitive dilatometry~\cite{Fukao2,Bauer1}. 
However, as shown in Fig.~2, $T_{\rm g}$ can be
successfully determined by capacitive  dilatometry in the case of
PS-DR1.
Furthermore, the thickness dependence of 
$T_{\rm g}$ in thin films of PS-DR1 can also be determined and is 
found to be consistent with that of unlabeled PS. From this 
result we can conclude that the rotational reorientation dynamics of DR1 
are an excellent sensor for the $\alpha$-process in thin films of
PS, and the results obtained in thin films of PS-DR1 can be compared
with those of unlabeled PS. 

\begin{figure}
\includegraphics[width=8cm]{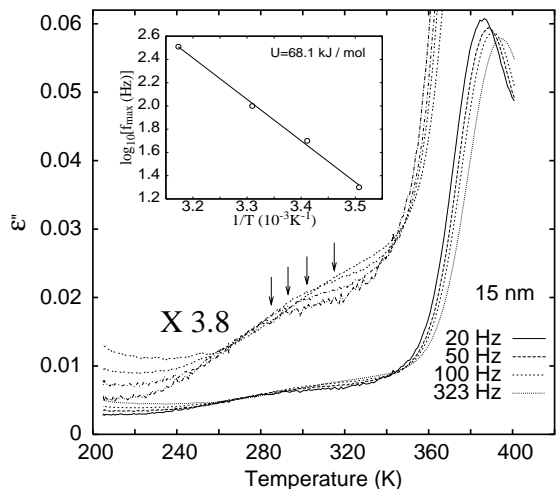}
\caption{\label{fig:11} The dielectric loss as a function of temperature
for different frequencies 20 Hz, 50 Hz, 100 Hz and 323 Hz for thin films of 
PS-DR1 with $d$ = 15 nm. The data below 360 K are magnified by 3.8 times.
The arrows indicate the location of the
 $\alpha_l$-process. The inset shows the Arrhenius plot for the
 $\alpha_l$-process.}
\end{figure}



From Fig. 3 it is observed that $T_{\rm g}$ decreases with decreasing 
film thickness for PS-DR1 in a similar manner as unlabeled PS.  The 
reduction in $T_{\rm g}$ with decreasing film thickness is believed to 
be related to a layer at the upper Al-electrode polymer interface with 
a reduced $T_{\rm g}$.  Thus, with decreasing film thickness, the layer 
with a reduced $T_{\rm g}$ contributes more to the average dynamics of 
the film, leading to a decrease in the average film $T_{\rm g}$.  

In a previous report by Fukao et al., it has been reported that the 
fragility decreases slightly with
decreasing film thickness in thin films of PS on the basis of 
the combined results of dielectric relaxation spectroscopy and 
thermal expansion spectroscopy~\cite{Fukao3}. 
In a recent paper, a 
slight decrease in the fragility index from 150 to 110 was also observed 
using dielectric
relaxation spectroscopy when the thickness was decreased from 286~nm to 
8.7~nm~\cite{Wubbenhorst2}.
The fragility index of PS-DR1 as a function of film thickness is shown 
in Table~II. Although there is a large error, the observed
results are qualitatively consistent with the previous results of unlabeled 
polystyrene.


The shapes of the dielectric loss and the distribution of
relaxation times of the $\alpha$-process for
thin films of PS-DR1 are shown in Figs.~8 and 9. The best fit parameters of
the HN equation are listed for thin films of PS-DR1 with various
film thickness at 390~K in Table~I. It is illustrated in Figs.~8 and 9 that
the distribution of the $\alpha$-relaxation times becomes broader with
decreasing film thickness,
which indicates that the thickness dependence of the distribution of the 
$\alpha$-relaxation times in PS-DR1 is the same as that of unlabeled PS.
The broadening of the distribution of the $\alpha$-relaxation times with
decreasing film thickness may result from a region at the upper
Al-polymer interface with dynamics different from bulk dynamics.  This
hypothesis will be tested in a forthcoming paper~\cite{Fukao6}. 
We note that the distribution of PS-DR1 is narrower than that of 
unlabeled PS at a fixed thickness:
$\beta_{KWW}$ = 0.53 for PS-DR1 with $d$ = 360 nm and 
$\beta_{KWW}$ = 0.435 for 
unlabeled PS with $d$ = 408 nm~\cite{Fukao2}. 
This may be related to the fact that the rotational 
reorientation dynamics of the DR1, which are coupled with the
$\alpha$-process, are observed by dielectric relaxation 
spectroscopy.

Figure~6 we shows that there is a decrease in the strength of
the $\alpha$-relaxation process with decreasing film thickness;
$\Delta\epsilon$ changes from 2.14 to 0.33 with decreasing film
thickness from 360~nm to 19~nm. This thickness dependence of 
$\Delta\epsilon$ is commonly observed in thin films of other
polymeric systems such polystyrene, poly(methyl methacrylate),
poly(vinyl acetate) and so on~\cite{Fukao4,Kremer2}. 
In a previous report by Fukao et al., a simple model for the decrease 
in $\Delta\epsilon$ 
in thin films was proposed and can be applied to thin films
of PS labeled with DR1. 
In the model, it is assumed that there is a motional unit in which $n$
dipole motions move or rotate cooperatively. In this case
$\Delta\epsilon$ is given by the following relation:
\begin{eqnarray}
\Delta\epsilon&=&\frac{N\mu^2}{3k_BT},
\end{eqnarray}
where N is the number of the motional units and is given by
$N_0=N\times n$, and $\mu$ is the total
strength of dipole moments included in a unit and is given by
$\mu=n\mu_0$.
Here, $\mu_0$ is the strength of a single dipole moment attached to
polymer chains, $N_0$ is the total number of dipole moments in the
system  and it is assumed that there is no correlation between 
the motional units. Using $N_0$ and $\mu_0$, we can rewrite
Eq.(9) as follows:
\begin{eqnarray}
\Delta\epsilon&=& n\frac{N_0\mu_0^2}{3k_BT}.
\end{eqnarray}
Therefore, if the number of dipole moments within the motional unit is
decreased with decreasing film thickness, the decrease in
$\Delta\epsilon$ in thinner films can be accounted for. The idea of
a decrease in the number of dipole moments moving cooperatively
is consistent with that based on the existence of a cooperatively
rearranging region (CRR)~\cite{Adam-Gibbs}.

In supported ultrathin films of unlabeled PS, it has been reported that
there is an additional relaxation process ($\alpha_l$-process) 
in addition to the 
$\alpha$-process~\cite{Fukao2,Wubbenhorst2}. The $\alpha_l$-process
is located at a lower temperature than that of the $\alpha$-process.
This process was assigned to
relaxation dynamics of the surface region in PS films, and 
has an Arrhenius type of temperature 
dependence with an activation energy of 71~kJ/mol~\cite{Wubbenhorst2}. 
A recent study investigating the relaxation processes of thin supported
polystyrene films using dielectric spectroscopy also observed an
additional relaxation process below $T_{\rm g}$  with an Arrhenius 
temperature dependence~\cite{Svanberg}. The activation energy of the additional
process was 15-25 kJ/mol.  However, the additional process was
attributed to a simple or primitive dynamical process that acts as a 
precursor to the glass transition in ultrathin free standing films.  
In the present study, an $\alpha_l$-process is observed in ultrathin 
PS-DR1 films ($d<$ 20 nm) at temperatures lower than that of the 
$\alpha$-process, as shown in Fig.~11.
%
From the investigation in Fig.~11, it is found that the  
$\alpha_l$-process of thin films of PS-DR1 can also be described by an
Arrhenius type of activation process with the activation energy 
$U$= 68$\pm$3~kJ/mol, which agrees very well with the value reported for the
unlabeled PS~\cite{Wubbenhorst2}.
At present, we cannot provide an unambiguous answer as to whether the 
additional relaxation process observed in the current study results from 
interfacial effects or a simple or primitive dynamical process.  
Studies are currently underway to determine whether or not it is an 
interfacial effect.

Here, it should be noted that the DR1 chromophore covalently attached
to the main chain in PS-DR1 is a bulky group. Therefore, there is a
possibility that the local structure of the amorphous PS chains 
deviates from that of the unlabeled PS and its deviation affects the
dynamics of the polymer chains observed by dielectric relaxation 
spectroscopy. 
However, we believe that
such deviation, if any, has almost no effect on the segmental motions of
PS-DR1, because the $\alpha$-dynamics
and its thickness dependence are consistent with those of unlabeled 
PS, as shown in our study. In addition, previous studies conducted using 
DR1 as the dye either doped or covalently attached to the polymer 
yielded average $\alpha$-relaxation times in agreement with those 
determined using other techniques.

\section{Summary}

We investigated the glass transition temperature and 
relaxation dynamics of the $\alpha$-process of thin films of
polystyrene labeled with a dye DR1 using dielectric relaxation measurements.
The results can be summarized as follows:
\begin{enumerate}
\item
The dielectric strength of DR1-labeled polystyrene is approximately 65 times 
as large as that of unlabeled polystyrene above the glass
transition, while there is almost no difference between them below the
glass transition. 
\item
The $T_{\rm g}$ of DR1-labeled
polystyrene can be determined well as a crossover temperature at which 
the temperature coefficient of the electric capacitance changes from the
value of the glassy state to that of the liquid state.
The $T_{\rm g}$ thus obtained decreases
with decreasing film thickness in a manner similar to that of
unlabeled polystyrene thin films.  
\item
As for the dielectric relaxation spectrum of the DR1-labeled polystyrene,
the $\alpha$-relaxation time becomes smaller and the distribution of the
$\alpha$-relaxation times becomes broader, as thickness decreases. 
\end{enumerate}

These results show
that there is a distinct contrast between the relaxation strength of
the $\alpha$-process of PS-DR1 and that of the unlabeled PS and that
thin films of DR1-labeled polystyrene are a suitable system
for investigating confinement effects of the glass transition
dynamics using dielectric relaxation spectroscopy.
Therefore, we expect that we will be able to observe the dynamics of the $\alpha$-process 
only from the labeled layer in a multi-layer system of PS-DR1 and 
unlabeled PS, and to obtain information on  
the dynamics of the $\alpha$-process at any position normal to the
film surface. We will report the results on such position dependent
measurements of the $\alpha$-dynamics in the near future~\cite{Fukao6}.\\

\section*{Acknowledgements}

This work was supported by the NSF-MRSEC Program at Northwestern University
(Grants DMR-0076097 and DMR-0520513), a Grant-in-Aid for Scientific Research
(B) (No.16340122) from Japan Society for the Promotion of Science, a DFI 
fellowship (R.D.P.) and a NSF EASPI fellowship (R.D.P.).

\end{document}